\documentclass[preprint,showpacs,preprintnumbers,amsmath,amssymb]{revtex4}

\usepackage{graphicx}
\usepackage{dcolumn}
\usepackage{bm}

\begin{document}


\title{The small mixing angle $\theta_{13}$ and the lepton asymmetry}

\author{Song-Haeng ~Lee} \email{shlee@phys.cau.ac.kr}
\author{Kim ~Siyeon} \email{siyeon@cau.ac.kr}

\affiliation{Department of Physics,
        Chung-Ang University, Seoul 156-756, Korea}
\date{May ~1, ~2005}
\begin{abstract}
We present the correlation of low energy CP phases, both Dirac and
Majorana, and the lepton asymmetry for the baryon asymmetry in the
universe, with a certain class of Yukawa matrices that consist of
two right-handed neutrinos whose mass ratio is about $2 \times
10^{-4}$ and include one texture zero in themselves. For cases in
which the amount of the lepton asymmetry $Y_L$ turns out to be
proportional to $\theta_{13}^2$, we consider the constraint
between two types of CP phases and the relation of $Y_L$ versus
the Jarlskog invariant or the amplitude of neutrinoless double
beta decay as $\theta_{13}$ varies.
\end{abstract}


\maketitle \thispagestyle{empty}

\newcommand{\cm}{\check{m}}
\newcommand{\yuk}{\mathcal{Y}}


\section{Introduction}
A study of neutrino masses is launched from the result which is
that two of three mixing angles in the lepton sector are large and
the other is small. Interpreting the atmospheric and solar
neutrino experiments \cite{atm}\cite{SNO} in terms of two-flavor
mixing, the mixing angle for the oscillation of atmospheric
neutrinos is understood to be maximal or nearly maximal: $
\sin^22\theta_{\rm{atm}} \simeq 1$, whereas the one for the
oscillation of solar neutrinos is not maximal but large: $
\sin^2\theta_{\rm{sol}} \simeq 0.3$. The masses of charged leptons
are the most precisely measured parameters of the fermions. The
data reads $m_e=0.51 ~\rm{MeV}, ~m_\mu=106 ~\rm{MeV}, ~m_\tau=1780
~\rm{MeV}$. Meanwhile, as for neutrinos, the existing data just
show that the neutrino mass squared differences which induce the
solar and atmospheric neutrino oscillations are $\Delta m_{sol}^2
\simeq \left( 7^{+10}_{-2} \right) \times 10^{-5}~\mbox{eV}^2$ and
$\Delta m^2_{atm}\simeq \left( 2.5^{+1.4}_{-0.9} \right) \times
10^{-3}~\mbox{eV}^2$, respectively. With SNO\cite{SNO} and
KamLAND\cite{Eguchi:2002dm}, data have narrowed down the possible
mass spectrum of neutrinos into two types, normal hierarchy
($m_1\lesssim m_2<m_3$) and inverse hierarchy ($m_3<m_1\lesssim
m_2$). If the experimental results $\Delta m_{sol}^2=m_2^2-m_1^2$
and $\Delta m_{atm}^2=|m_3^2-m_2^2|$  are accommodated to the
masses of normal hierarchy, one can obtain the following relations
for mass ratio,
\begin{eqnarray}
    \frac{m_2}{m_3} \approx
    \sqrt{\frac{\Delta m_{sol}^2}{\Delta m_{atm}^2}}=
    \left(1.7^{+1.7}_{-.6}\right) \times 10^{-1},
    \label{nhratio}
\end{eqnarray}
assuming $m_1$ is strongly restricted to be smaller than $m_2$ by
the order of magnitude, rather than $m_1\lesssim m_2$.

The upper bound $ \sin^22\theta_{13} < 0.2 $ was obtained from the
Chooz experiment\cite{chooz}. At the present stage in neutrino
physics, the yet-unknown $\theta_{13}$ is considered as the key
element to shed light on the feasibility of CP violation and the
mass hierarchy. A future long-baseline experiment proposed
currently intends to improve the sensitivity up to $
\sin^22\theta_{13} < 0.05 $ in combination with a reactor
experiment, and further $ \sin^22\theta_{13} < 0.02 $ as beam
rates get enhanced\cite{Diwan:2003bp}.

Including data from SNO\cite{SNO} and KamLand\cite{Eguchi:2002dm},
the range of the magnitude of the MNS mixing
matrix\cite{Maki:1962mu} at the 90\% confidence level is given by
\cite{Guo:2002ei},
    \begin{eqnarray}
    |U| = \left(
        \begin{array}{ccc}
        0.79-0.86 & 0.50-0.61 & 0-0.16 \\
        0.24-0.52 & 0.44-0.69 & 0.63-0.79 \\
        0.26-0.52 & 0.47-0.71 & 0.60-0.77
        \end{array}\right),
    \label{mixnum}
    \end{eqnarray}
where the unitary mixing matrix is defined via
$\nu_{a}=\sum^{3}_{j=1}U_{aj}\nu_{j}~(a=e,\mu,\tau)$, with a
flavor eigenstate  $\nu_a$ and a mass eigenstate $\nu_j$. It can
be readily recognized that the central values of elements in the
mixing matrix in Eq.(\ref{mixnum}) are pointing likely numbers,
$\sin\theta_{sol}=\frac{1}{\sqrt{3}}$ and
$\sin\theta_{atm}=\frac{1}{\sqrt{2}}$
\cite{tribi}\cite{Chang:2004wy}.

The purpose of this work is to find the relation between dirac
phase and majorana phase eligible to the compatibility of the high
energy lepton asymmetry to explain the baryon asymmetry in
universe and the smallness of $\theta_{13}$. The former was
estimated via Cosmic Microwave Background and the later is about
to be explored in near-future experiments with higher
precision\cite{Diwan:2003bp}. As to be shown, when both types of
CP violating phases, Dirac and Majorana, are considered, the
lepton asymmetry, in general, does not approach to zero as
$\theta_{13}$ does so. The canonical seesaw mechanism with two
right-handed neutrinos will be exploited for the connection of the
phenomena in different scales. When we utilize the seesaw
mechanism in bottom-up direction, the input parameters are set up
from the observed results, and the prediction from a model
constructed in Ref.\cite{Siyeon:2004wb} such as
    \begin{eqnarray}
        \sqrt{2}\sin\theta_{13}=\frac{2}{3}\frac{m_2}{m_3}
        \label{discrete1}\\
        \frac{M_1}{M_2} \approx 2\times 10^{-4}.
        \label{discrete2}
    \end{eqnarray}
If one takes the estimation for $m_2/m_3$ in Eq.(\ref{nhratio}),
the prediction in Eq.(\ref{discrete1}) corresponds to
$\sin^22\theta_{13}=0.0255$, which can be tested in reactor and
long-baseline experiment with enhanced sensitivity. Furthermore, a
somehow smaller value of $\sin\theta_{13}$ by a factor 0.8 or 0.6,
which can be tested with higher resolution in future experiments,
will be analyzed together. The ratio in Eq.(\ref{discrete2}) was
obtained by the same breaking parameter from $U(1) \times Z_3
\times Z_2$ flavor symmetry by the same mechanism, as the small
angle in Eq.(\ref{discrete1}) was obtained. So, the test of the
model giving rise to a certain ratio of $M_1/M_2$ in
Eq.(\ref{discrete2}) can be done, once $\sin{\theta_{13}}$ is
measured in experiment. The testability motivates us to take them
as inputs in this analysis.

The following contents are organized in order. In Section II, the
mass matrix of light neutrinos is given in terms of low energy
measurable parameters. In Section III, possible neutrino Yukawa
matrices with two right-handed neutrinos are derived through
bottom-up seesaw mechanism. The magnitude of CP asymmetry in
decays of heavy Majorana neutrinos will be estimated as well. In
Section IV, the lepton asymmetry to produce the baryon asymmetry
in the universe via sphaleron process is evaluated from the CP
asymmetry and the dilution factor which are parameterized in terms
of low energy CP phases and the small mixing angle $\theta_{13}$.
With respect to different values of $\theta_{13}$'s, the
$\delta$-$\varphi$ space is tested for the amount of lepton
asymmetry dictated by the observed baryon asymmetry. For a
particular case in which the CP asymmetry in decays of heavy
neutrinos or the dilution factor has strong dependence on the
mixing angle $\theta_{13}$, some consideration on the correlation
of the lepton asymmetry to the size of low energy CP violation and
that to the amplitude of neutrinoless double beta decay follows as
a remark in the last section.

\section{Masses and mixing angles of light neutrinos}

In general, a unitary mixing matrix for 3 generations of neutrinos
is given by
    \begin{eqnarray}
    \tilde{U} &=& R\left(\theta_{23}\right)
          R\left(\theta_{13},\delta\right)
          R\left(\theta_{12}\right)P(\varphi, \varphi')
    \label{fulltrans}
    \end{eqnarray}
where $R$'s are rotations with three angles and a Dirac phase
$\delta$ and the $P=Diag\left(1,e^{i\varphi/2},e^{i\varphi'/2}
\right)$ with Majorana phases $\varphi$ and $\varphi'$ is a
diagonal phase transformation. The mass matrix of light neutrinos
is given by $M_{\nu} = \tilde{U} Diag(m_1,m_2,m_3) \tilde{U}^T$,
where $m_1,m_2,m_3$ are the real positive masses of light
neutrinos. Or the Majorana phases can be embedded in the diagonal
mass matrix such that
    \begin{equation}
    M_{\nu} = U Diag(m_1,\cm_2,\cm_3) U^T,
    \label{umu}
    \end{equation}
where $U \equiv \tilde{U}P^{-1}$ and $\cm_2 \equiv
m_2e^{i\varphi}$ and $\cm_3 \equiv m_3e^{i\varphi'}$. If
transformation matrix $U$ in the standard parametrization for CKM
\cite{Eidelman:2004wy} is
\begin{widetext}
    \begin{eqnarray}
        \left(\begin{array}{ccc}
            c_{12}c_{13} & s_{12}c_{13} & s_{13}e^{-i\delta}\\
            -s_{12}c_{23}-c_{12}s_{23}s_{13}e^{i\delta} &
            c_{12}c_{23}-s_{12}s_{23}s_{13}e^{i\delta} &
            s_{23}c_{13} \\
            s_{12}s_{23}-c_{12}c_{23}s_{13}e^{i\delta} &
            -c_{12}s_{23}-s_{12}c_{23}s_{13}e^{i\delta} &
            c_{23}c_{13}
        \end{array}\right),\label{ckm}
    \end{eqnarray}
\end{widetext}
where $s_{ij}$ and $c_{ij}$ denotes $\sin{\theta_{ij}}$ and
$\cos{\theta_{ij}}$ with the mixing angle $\theta_{ij}$ between
$i$-th and $j$-th generations, and if
$s_{12}=\frac{1}{\sqrt{3}}(1+\sigma),\hspace{3pt} \sigma \ll 1$
and $s_{23}=\frac{1}{\sqrt{2}}, s_{13}\ll 1$ are adopted for the
angles in the transformation $U$, the matrix $M_\nu$ can be
expressed as
\begin{widetext}
\begin{eqnarray}
    M_{\nu} \approx
    \frac{m_1}{6}\left(\begin{array}{ccc}
    4 & -2 & -2 \\ -2 & 1 & 1 \\ -2 & 1 & 1
    \end{array}\right) +
        \frac{\cm_2}{3}\left(\begin{array}{ccc}
        1+2\sigma & 1+\frac{1}{2}\sigma & 1+\frac{1}{2}\sigma \\
        & 1-\sigma & 1-\sigma \\
        &  & 1-\sigma
        \end{array}\right) +
            \frac{\cm_3}{2}\left( \begin{array}{ccc}
            2\vartheta^2 & -\sqrt{2}\vartheta & \sqrt{2}\vartheta \\
            & 1-\vartheta^2 & -1+\vartheta^2 \\
            &  & 1-\vartheta^2 \end{array} \right),
    \label{mass1}
\end{eqnarray}
\end{widetext}
where $\vartheta=s_{13}e^{-i\delta}$ with a Dirac phase $\delta$.

Inspired from the form of mass in Eq.(\ref{mass1}), an ansatz for
mass matrix of neutrinos is introduced by adopting the leading
contributions such that,
\begin{widetext}
\begin{eqnarray}
    \frac{M_{\nu}}{m^*}
    &=& \left(\begin{array}{rrr}
    0 & 0 & 0 \\ 0 & 1 & -1 \\ 0 & -1 & 1
    \end{array}\right) +
        \rho e^{i\varphi}
        \left( \begin{array}{ccc}
        1 & 1 & 1 \\
        1 & 1 & 1 \\
        1 & 1 & 1 \end{array} \right) +
            \theta e^{-i\delta}
            \left( \begin{array}{ccc}
            0 & -1 & 1 \\
            -1 & 0 & 0 \\
            1 & 0 & 0 \end{array} \right),
    \label{morder}
\end{eqnarray}
\end{widetext}
which is divided by $m^*$ of mass dimension. It was shown that the
above structure of neutrino mass could be derived using a flavor
symmetry\cite{Siyeon:2004wb}. We will take a value $\rho = 0.11$
that corresponds to $(2/3)$ times $\sqrt{\Delta m_{sol}^2/\Delta
m_{atm}^2}$ given in Eq.(\ref{nhratio}). Regarding $\theta$ that
corresponds to $\sqrt{2}\sin\theta_{13}$, we take its upper bound
to be $\rho$, such as $\theta \le \rho$. When the $\theta$ varies
from $0$ to $\rho$, the dimensionless matrix in Eq.(\ref{morder})
gives rise to the magnitude of transformation with such ranges as
\begin{eqnarray}
    \left(
        \begin{array}{ccc}
        0.816 - 0.825 \quad &
        0.577 - 0.560 \quad &
        0 - 0.0821  \\
        0.408 - 0.352 \quad &
        0.577 - 0.622 \quad &
        0.707 - 0.700 \\
        0.408 - 0.443 \quad &
        0.577 - 0.548 \quad &
        0.707 - 0.710
        \end{array}\right),
    \label{range}
\end{eqnarray}
while the mass ratios take ranges as $m_2/m_3 = 0.165 - 0.162$ and
$m_1/m_3 = 1 \times 10^{-10} - 4 \times 10^{-3}$. The ranges
should be read as a value at $\theta=0$ to that at $\theta= \rho$.
Thus, it is clear that the smallest mass eigenvalue can be taken
as zero and accordingly only a single relative Majorana phase as
in Eq.(\ref{morder}) can be chosen without loss of generality.

When $\theta = \rho$, the Jarlskog invariant
$\mathcal{J}=Im[U_{e1}U_{\tau3}U_{\tau1}^*U_{e3}^*]$ of the
transformation of the mass matrix in Eq.(\ref{morder}) can be
evaluated using Eq.(\ref{ckm}) and Eq.(\ref{range}). The
$s_{12}s_{23}$, the first term in $U_{\tau1}$ is simply
$U_{\tau1}$ if $s_{13}=0$. So, by taking the sizes of $U_{e1},
U_{\tau3}, U_{e3}$ at $\theta=\rho$ and that of $U_{\tau1}$ at
$\theta=0$, we can obtain the magnitude of CP violation as
\begin{eqnarray}
    \mathcal{J}=
    |U_{e1}U_{e3}U_{\tau3}|_{\theta=\rho}|U_{\tau1}|_{\theta=0}
    \sin\delta,
\end{eqnarray}
whose measurement is possible from neutrino oscillations. Although
Majorana CP violating phases are not detectable through neutrino
oscillations, they may affect the amplitude of neutrinoless double
beta decay $<m_{ee}>\equiv|\sum^3_{i=1}U_{ei}^2m_ie^{i\varphi_i}|$
where $\varphi_i$ denotes the Majorana phases. The amplitude with
a single Majorana phase from the ansatz in Eq.(\ref{morder}) and
the numerical values in Eq.(\ref{range}) is
    \begin{eqnarray}
    <m_{ee}>&=& m_3
    \sqrt{\rho^2|U_{e2}|^2+|U_{e3}|^2+2\rho|U_{e2}U_{e3}|\cos(\varphi-2\delta)},
    \end{eqnarray}
if one of the masses is considered as zero so that the number of
physical Majorana phases is one.

In the following sections, the parameters are set up such that two
mass eigenvalues and two large mixing angles are fixed based on
experimental data while two CP phases are still free with respect
to light neutrinos. The small angle $\theta_{13}$ and the ratio
$M_1/M_2$ of heavy right neutrinos are taken as in
Eq.(\ref{discrete1}) and Eq.(\ref{discrete2}) predicted from a
particular model \cite{Siyeon:2004wb}, while $M_2$ is chosen to be
of the scale of Grand Unified Theory. As for $\theta_{13}$, the
value for $\theta= 0.6 \rho$ involves the lepton asymmetry curve
of which extremum is about that of the observed baryon asymmetry
to be shown in the last section, so that the value showing such
aspect will be considered as well as the value of
Eq.(\ref{discrete1}) for comparison.

\section{Seesaw mechanism and neutrino Yukawa matrices}

Neutrino mass models with one zero mass involved in three active
neutrinos can be generated naturally from the seesaw mechanism
with two right-handed neutrinos. In the basis mass matrix $M_R$ of
right-handed neutrinos $N_R=(N_1, N_2)$ is diagonal, a model is
given in terms of Yukawa interactions of leptons and lepton number
violating Majorana mass $M_R$ of right-handed neutrinos:
\begin{eqnarray}
    - \mathcal{L} = H \yuk_\ell L_e \bar{e}_R
    + H \yuk_\nu L_e \bar{N}_R + \frac{1}{2} M_R N_R N_R,
\end{eqnarray}
which consists of a $3 \times 3$ matrix $\yuk_\ell$, a $3 \times
2$ matrix $\yuk_\nu$ a $2 \times 2$ matrix $M_R$, with two
right-handed neutrinos. When Yukawa interaction of neutrinos is
introduced, the light masses are derived through the seesaw
mechanism\cite{Gell-Mann:vs}, $M_\nu = - v^2 \yuk_\nu M_R^{-1}
\yuk_\nu^T$ in top-down approach. On the other hand, the matrix
$\yuk_\nu$ is found as the solution to the seesaw mechanism in
bottom-up approach once one launches the analysis with the light
neutrino masses $M_\nu$. Let $M_1$ and $M_2$ be the masses of
right-handed neutrinos and $M_{ij}$ the elements of the matrix
$M_\nu$. If Yukawa matrix is given as,
    \begin{equation}
        \yuk_\nu=\left(
        \begin{array}{cc}
        \sqrt{\mu} a_1 & b_1 \\
        \sqrt{\mu} a_2 & b_2 \\
        \sqrt{\mu} a_3 & b_3
        \end{array} \right),
    \label{dirac}
    \end{equation}
where $\sqrt{\mu}\equiv \sqrt{M_1/M_2}$. The elements $a_j$'s and
$b_j$'s are distinguished such that only one of the $b_j$'s is
zero, and then a matrix $\yuk_\nu=(\sqrt{\mu}b_j~ a_j)$ also will
be used when a texture zero is placed in the first column. One can
obtain its entries in terms of dimensionless $\tilde{M}_{jk}\equiv
M_{jk}/m^*$ with $j,k=1,2,3$,
\begin{widetext}
    \begin{eqnarray}
        && a_1 = \sqrt{\tilde{M}_{11} - b_1^2},\mbox{   or   }
        b_1 = \sqrt{\tilde{M}_{11} - a_1^2}, \nonumber \\
        && a_i = \frac{1}{\tilde{M}_{11}} \left[ a_1 \tilde{M}_{1i} -
        \sigma_i b_1 \sqrt{\tilde{M}_{11}\tilde{M}_{ii}-\tilde{M}_{1i}^2} \right],
        \nonumber \\
            && b_i = \frac{1}{\tilde{M}_{11}} \left[ b_1 \tilde{M}_{1i} +
            \sigma_i a_1 \sqrt{\tilde{M}_{11}\tilde{M}_{ii}-\tilde{M}_{1i}^2}\right],
            \label{key} \\
            && \tilde{M}_{11}\tilde{M}_{23} =  \left[\tilde{M}_{12}\tilde{M}_{13} +
            \sigma_2\sigma_3
            \sqrt{\left(\tilde{M}_{11}\tilde{M}_{22}-\tilde{M}_{12}^2\right)
            \left(\tilde{M}_{11}\tilde{M}_{33}-\tilde{M}_{13}^2\right)} \right].\nonumber
    \end{eqnarray}
\end{widetext}
where the $i$ is 2 or 3, the $\sigma_i$ is a sign $\pm$, and the
sign of $a_1$ is fixed as positive. However, the relative sign of
$a_1$ to $b_1$ is still undetermined. The above expression derived
in Ref.\cite{Barger:2003gt} is equivalent to the neutrino mass
matrix in the following way:
    \begin{eqnarray}
        M_\nu=\frac{v^2}{M_2}
    \left(\begin{array}{ccc}
        a_1^2+b_1^2 & a_1a_2+b_1b_2 & a_1a_3+b_1b_3 \\
        a_1a_2+b_1b_2 & a_2^2+b_2^2 & a_2a_3+b_2b_3 \\
        a_1a_3+b_1b_3 & a_2a_3+b_2b_3 & a_3^2+b_3^2
    \end{array}\right)
    \label{abc}
    \end{eqnarray}
It is clear that only 5 out of 6 parameters $(a_1, b_1, a_i, b_i)$
can be specified in terms of the elements of $M_\nu$, since 6
independent elements of the symmetric matrix are related by the
zero determinant. There are various ways to decrease the number of
parameters in a Yukawa matrix, posing a texture zero or posing
equalities between elements for the matrix texture. It is known
that texture zeros or equalities among matrix entries can be
generated by imposing additional symmetries to the theory. In this
work, we focus on the types of Yukawa matrices with a texture
zero.

If one takes the following matrix introduced in Eq.(\ref{morder})
\begin{eqnarray}
    \left(\begin{array}{ccc}
        \rho e^{i\varphi} & \quad \rho e^{i\varphi}-\theta e^{-i\delta} &
            \quad \rho e^{i\varphi}+\theta e^{-i\delta} \\
        \rho e^{i\varphi}-\theta e^{-i\delta} & \quad \rho e^{i\varphi}+1 &
            \quad \rho e^{i\varphi}-1 \\
        \rho e^{i\varphi}+\theta e^{-i\delta} & \quad \rho e^{i\varphi}-1 &
            \quad \rho e^{i\varphi}+1
    \end{array}\right)
    \label{lightinput}
\end{eqnarray}
for $\tilde{M}_{jk}$ used in Eq.(\ref{key}), it will give rise to
the 6 possible matrices for $\yuk_\nu$ in Eq.(\ref{dirac}) which
consist of the three with one of Yukawa couplings of $N_2$ absent;
\begin{widetext}
    \begin{eqnarray}
        && \yuk\hspace{2pt}[y_{12}=0]=\frac{1}{\sqrt{\rho e^{i\varphi}}}\left(
        \begin{array}{ll}
        \sqrt{\mu}\rho e^{i\varphi} & \quad 0 \\
        \sqrt{\mu}(\rho e^{i\varphi}-\theta e^{-i\delta}) &
        \quad \sqrt{\rho e^{i\varphi}
        (\rho e^{i\varphi}+1)-(\rho e^{i\varphi}-\theta e^{-i\delta})^2} \\
        \sqrt{\mu}(\rho e^{i\varphi}+\theta e^{-i\delta}) &
        \quad -\sqrt{\rho e^{i\varphi}(\rho e^{i\varphi}+1)-(\rho e^{i\varphi}+\theta e^{-i\delta})^2}
        \end{array} \right), \nonumber \\ \nonumber \\
            && \yuk\hspace{2pt}[y_{22}=0]=\frac{1}{\sqrt{\rho e^{i\varphi}+1}}\left(
            \begin{array}{ll}
            \sqrt{\mu}(\rho e^{i\varphi}-\theta e^{-i\delta}) &
            \quad -\sqrt{\rho e^{i\varphi}
            (\rho e^{i\varphi}+1)-(\rho e^{i\varphi}-\theta e^{-i\delta})^2} \\
            \sqrt{\mu}(\rho e^{i\varphi}+1) & \quad 0 \\
            \sqrt{\mu}(\rho e^{i\varphi}-1) &
            \quad \sqrt{(\rho e^{i\varphi}+1)^2-(\rho e^{i\varphi}-1)^2}
            \end{array} \right), \label{onezero} \\ \nonumber \\
        && \yuk\hspace{2pt}[y_{32}=0]=\frac{1}{\sqrt{\rho e^{i\varphi}+1}}\left(
        \begin{array}{ll}
        \sqrt{\mu}(\rho e^{i\varphi}+\theta e^{-i\delta}) &
        \quad \sqrt{\rho e^{i\varphi}
        (\rho e^{i\varphi}+1)-(\rho e^{i\varphi}+\theta e^{-i\delta})^2} \\
        \sqrt{\mu}(\rho e^{i\varphi}-1) & \quad -\sqrt{(\rho e^{i\varphi}+1)^2-
        (\rho e^{i\varphi}-1)^2} \\
        \sqrt{\mu}(\rho e^{i\varphi}+1) & \quad 0
        \end{array} \right), \nonumber
    \end{eqnarray}
\end{widetext}
and three more matrices, $\yuk\hspace{2pt}[y_{11}=0],
\hspace{2pt}\yuk\hspace{2pt}[y_{21}=0]$, and
$\yuk\hspace{2pt}[y_{31}=0]$, which make the counterparts to the
above three sets by exchanging their columns, that is, when one of
Yukawa couplings of $N_1$ is absent.

Once we have the neutrino Yukawa matrix, we can calculate the
magnitude of CP asymmetry $\epsilon_i$ in decays of heavy Majorana
neutrinos \cite{Kolb:qa,Luty:un},
\begin{eqnarray}
\epsilon_i
        &=&\frac{\Gamma (N_i \to \ell H)
            - \Gamma (N_i \to \bar{\ell} H^*)}
        {\Gamma (N_i \to \ell H)
            + \Gamma (N_i \to \bar{\ell} H^*)},
\label{aacp}
\end{eqnarray} %
where $i$ denotes a generation. When one of two generations of
right neutrinos has a mass far below that for the other
generation, i.e., $M_1 \ll M_2$, the $\epsilon_i$ in
Eq.(\ref{aacp}) can be replaced by just $\epsilon_1$ obtained from
the decay of $M_1$ \cite{Kolb:qa,Luty:un},
\begin{eqnarray}
    \epsilon_1 &=& \frac{1}{8\pi}
            \frac{{\rm Im}\left[(\yuk_\nu^\dagger \yuk_\nu)_{12}^2\right]
            } {(\yuk_\nu^\dagger \yuk_\nu)_{11}}
            f\left(\frac{M_2}{M_1}\right)\;, \label{cp1}
\end{eqnarray}
where $f\left(M_2/M_1\right)$ represents loop contribution to the
decay width from vertex and self energy and is given by
\begin{equation}
f(x) = x\left[1-(1+x^2)\ln \frac{1+x^2}{x^2} +
\frac{1}{1-x^2}\right]
\end{equation}
for the Standard Model. For a large value of $x$, the leading
order of $f(x)$ is of $3x^{-1}/2$. It is convenient to consider
separately the factor that depends on Yukawa matrix in
$\epsilon_1$ in Eq.(\ref{cp1}) at this stage.
    \begin{eqnarray}
        && \frac{{\rm Im}\left[(\yuk_\nu^\dagger \yuk_\nu)_{12}^2\right]}
            {(\yuk_\nu^\dagger \yuk_\nu)_{11}} =
        \frac{Im \left[( y_{11}^* y_{12} + y_{21}^* y_{22} + y_{31}^* y_{32})^2\right]}
        {|y_{11}|^2 + |y_{21}|^2 +|y_{31}|^2 }\equiv \Delta_1,
        \label{delta1} \\
            && \epsilon_1 = \frac{3}{16\pi}\Delta_1 \frac{M_1}{M_2},
            \label{finalCP}
    \end{eqnarray}
for $M_1 \ll M_2$, where $a$'s and $b$'s are defined in Eq.(11).
The ratio $M_1/M_2 \sim 2 \times 10^{-4}$ in Eq.(\ref{discrete2})
will be used for the evaluation of the CP asymmetry.

From a number of types of matrices with a texture zero derived in
Eq.(\ref{onezero}), 6 different values of $\Delta_1$'s can be
evaluated and expressed using the following polynomials:
\begin{eqnarray}
    && \mathcal{A}_1(\rho^0),\mathcal{A}_2(\rho^0) \equiv
        2+3\rho^2 \mp 2\rho\theta\cos{(\varphi+\delta)} +\theta^2 \label{A1A2}\\
    && \mathcal{B}_1(\rho),\mathcal{B}_2(\rho) \equiv \sqrt{\rho^2 \pm 4\rho^2 \theta
            \cos{\delta}-2\rho\theta^2\cos{(\varphi+2\delta)}
            +4\rho^2\theta^2 \mp
            4\rho\theta^3\cos{(\varphi+\delta)}+\theta^4},
            \label{B1B2}
\end{eqnarray}
where $\mathcal{A}_1(\rho^0)$ in Eq.(\ref{A1A2}) denotes the terms
with the upper sign and indicates that the leading term in
$\mathcal{A}_1$ is of order $\rho^0$. So does $\mathcal{A}_2$ with
lower sign. Likewise, $\mathcal{B}_1$ of leading order $\rho$ in
Eq.(\ref{B1B2}) denotes the terms with the upper sign inside the
square root. If $y_{11}$ and $y_{12}$ is zero, one can obtain the
following expression,
\begin{widetext}
\begin{eqnarray}
    && |a_1|^2+|a_2|^2+|a_3|^2 = \rho^{-1}(3\rho^2 +2\theta^2),\label{asum1} \\
    && |b_1|^2+|b_2|^2+|b_3|^2 = \rho^{-1}
        \{\mathcal{B}_1(\rho) +
        \mathcal{B}_2(\rho)\} \label{bsum1} \\
    && Im[(a_1^*b_1+a_2^*b_2+a_3^*b_3)^2]=C_0 + C_1\theta + C_2\theta^2 +
        \mathcal{O}(\theta^3), \label{imaginary1}\\
        && \qquad C_0 = 0, \nonumber \\
        && \qquad C_1 = 0,\nonumber\\
        && \qquad C_2 = 4\rho^{-1}\sin{(\varphi+2\delta)}
        +\mathcal{O}(\rho).\nonumber
\end{eqnarray}
\end{widetext}
If one of $y_{j2}$'s is zero, $\sum|y_{j1}|^2=\sum|a_j|^2,~
\sum|y_{j2}|^2=\sum|b_j|^2,~Im[(y_{j1}^*y_{j2})^2]=Im[(a_j^*b_j)^2]$,
while, if one of $y_{j1}$'s is zero, $\sum|y_{j1}|^2=\sum|b_j|^2,~
\sum|y_{j2}|^2=\sum|a_j|^2,~Im[(y_{j1}^*y_{j2})^2]=-Im[(a_j^*b_j)^2]$.
Thus, the CP asymmetry for $y_{11}=0$ or $y_{12}=0$ is
proportional to
\begin{widetext}
\begin{eqnarray}
    && \Delta_1[y_{11},y_{12}=0] =
        \left[ \frac{1}{\mathcal{B}_1(\rho)+\mathcal{B}_2(\rho)},
        \frac{-1}{3\rho^2+\theta^2} \right]
        \{ 4\theta^2\sin{(\varphi+2\delta)}+\mathcal{O}(\rho^2\theta^2)
        \},
        \label{delta11}
\end{eqnarray}
\end{widetext}
respectively. The approximation in Eq.(\ref{delta11}) is valid
unless $\pi-\rho < \varphi < \pi+\rho$.

When $y_{21}$ or $y_{31}$ is zero or when $y_{22}$ or $y_{32}$ is
zero can be considered in terms of the following expressions;
\begin{widetext}
\begin{eqnarray}
    && |a_1|^2+|a_2|^2+|a_3|^2 = \sqrt{\rho^2+2\rho\cos{\varphi}+1}^{-1}
        \Big\{\begin{array}{l}
            \mathcal{A}_1(\rho^0)
            \quad for \quad [b_2=0] \\
            \mathcal{A}_2(\rho^0)
            \quad for \quad [b_3=0]
            \end{array} \label{asum2} \\
    && |b_1|^2+|b_2|^2+|b_3|^2
            = \sqrt{\rho^2+2\rho\cos{\varphi}+1}^{-1} \left( 4\rho +
            \Big\{\begin{array}{l}
            \mathcal{B}_1(\rho)
            \quad for \quad [b_2=0] \\
            \mathcal{B}_2(\rho)
            \quad for \quad [b_3=0]
            \end{array} \right) \label{bsum2}\\
    && Im[(a_1^*b_1+a_2^*b_2+a_3^*b_3)^2]
        = (\rho^2+2\rho\cos{\varphi}+1)^{-1}
        \big\{C_0 + C_1\theta + C_2\theta^2 + \mathcal{O}(\theta^3) \big\},
        \label{imaginary2} \\
        && \qquad C_0 = 4\rho\sin{\varphi} -4\rho^2\sin(2\varphi)+ \mathcal{O}(\rho^5),\nonumber\\
        && \qquad C_1 = \mp 4\rho\sin{(\delta+\varphi)} +\mathcal{O}(\rho^2),\nonumber\\
        && \qquad C_2 = \mathcal{O}(\rho),
\end{eqnarray}
\end{widetext}
where $C_1$ takes $-$ sign if $b_2=0$, and $+$ sign if $b_3=0$.
Thus, the CP asymmetry is proportional to
\begin{widetext}
\begin{eqnarray}
    && \Delta_1 \left[ y_{21},y_{31}=0 \right] =
        \frac{\sqrt{\rho^2+2\rho\sin{\varphi}+1}}
        {4\rho+\left[ \mathcal{B}_1(\rho),\mathcal{B}_2(\rho) \right]}\{
        C_0(\rho)+C_1(\rho)\theta+C_2(\rho)\theta^2 \},\label{delta21}\\
    && \Delta_1 \left[ y_{22},y_{32}=0 \right] =
        -\frac{\sqrt{\rho^2+2\rho\sin{\varphi}+1}}
        {[\mathcal{A}_1(\rho^0),\mathcal{A}_2(\rho^0)]}\{
        C_0(\rho)+C_1(\rho)\theta+C_2(\rho)\theta^2 \}.
        \label{delta22}
\end{eqnarray}
\end{widetext}
The CP asymmetry in Eq.(\ref{finalCP}) is clearly parameterized by
the ratio of heavy neutrino masses and $\Delta_1$ expressed fully
by low-energy observables. Without $\phi$, the CP asymmetry
depends on $\theta_{13}^2\sin(2\delta)$ in the leading order if
$y_{11}$ or $y_{12}=0$, and $\theta_{13}\sin\delta$ otherwise. On
the other hand, the CP asymmetry without $\theta_{13}$ depends on
$\sin\varphi$ as shown in Ref.\cite{Chang:2004wy}. A number of
models were discussed about in focus of the connection of
measurable CP violations in low energy and the CP asymmetry for
the leptogenesis\cite{Endoh:2002wm}.

\section{Lepton asymmetry}

The baryon density of our universe $\Omega_B h^2 = 0.0224 \pm
0.0009$ implied by WMAP(Wilkinson Microwave Anisotropy Probe) data
indicates the observed baryon asymmetry in the
Universe\cite{WMAP1},
    \begin{equation}
    \eta_B^{CMB}=
    \frac{n_B-n_{\bar{B}}}{n_\gamma}=
    \left(6.5^{+0.4}_{-0.3}\right) \times 10^{-10},
    \label{baryon}
    \end{equation}
where $n_B, n_{\bar{B}}$ and $n_\gamma$ are number density of
baryon, anti-baryon and photon, respectively. The leptogenesis
\cite{lep} has become a compelling theory to explain the observed
baryon asymmetry in the universe, due to increasing reliance on
the seesaw mechanism from experiments. The baryon asymmetry
Eq.(\ref{baryon}) can be rephrased
    \begin{equation}
    Y_B = \frac{n_B-n_{\bar{B}}}{s} \simeq
    \left(8.8 - 9.8 \right) \times 10^{-11}. \label{cosm}
    \end{equation}
The $n_{\gamma}$ is the photon number density and the $s$ is
entropy density so that the number density with respect to a
co-moving volume element is taken into account. The baryon
asymmetry produced through sphaleron process is related to the
lepton asymmetry \cite{Harvey:1990qw} by $Y_B = \frac{a}{a-1} Y_L$
with $ a \equiv (8 N_F + 4 N_H) / (22 N_F + 13 N_H)$, for example,
$a=28/79$ for the Standard Model(SM) with three generations of
fermions and a single Higgs doublet, $N_F = 3, N_H = 1$. The
generation of a lepton asymmetry requires the CP-asymmetry and
out-of-equilibrium condition. The $Y_L$ is explicitly
parameterized by two factors, $\epsilon$, the size of CP
asymmetry, and $\kappa$, the dilution factor from washout process.
\begin{eqnarray}
Y_L = \frac{(n_L - n_{\overline{L}})}{s} = \kappa
\frac{\epsilon_i}{ g^*} \label{aalepto}
\end{eqnarray}
where $g^*\simeq 110$ is the number of relativistic degree of
freedom.

The $\kappa$ in Eq.(\ref{aalepto}) is determined by solving the
full Boltzmann equations. The $\kappa$ can be simply parameterized
in terms of $K$ defined as the ratio of $\Gamma_1$ the tree-level
decay width of $\nu_{R1}$ to $H$ the Hubble parameter at
temperature $M_1$, where $K\equiv \Gamma_1 / H<1$ describes
processes out of thermal equilibrium and $\kappa<1$ describes
washout effect\cite{Harvey:1990qw},
\begin{eqnarray}
    \kappa = \frac{0.3}{K \left(\ln K \right)^{0.6}}
    &\rm{for}& 10 \lesssim K \lesssim 10^6, \label{largek} \\
        \kappa = \frac{1}{2 \sqrt{K^2+9}}
        &\rm{for}& 0 \lesssim K \lesssim 10. \label{smallk}
\end{eqnarray}
The decay width of $N_1$ by the Yukawa interaction at tree level
and Hubble parameter in terms of temperature $T$ and the Planck
scale $M_{pl}$ are $\Gamma_1 =  (\yuk_\nu^\dagger \yuk_\nu)_{11}
M_1 / (8 \pi) $ and $H = 1.66 g^{1/2}_* T^2 / M_{pl}$,
respectively. At temperature $T = M_1$, the ratio $K$ is
    \begin{eqnarray}
    K = \frac{M_{pl}}{1.66 \sqrt{g^*}(8 \pi)}
    \frac{(\yuk_\nu^\dagger \yuk_\nu)_{11}}{M_1},
    \end{eqnarray}
which reduces to, since $(\yuk_\nu^\dagger
\yuk_\nu)_{11}=\mu\sum|a_i|^2$ in Eq.(\ref{dirac}),
    \begin{eqnarray}
    K = \frac{M_{pl}/M_2}{1.66 \sqrt{g^*}(8 \pi)} \sum|a_i|^2,
    \label{kay}
    \end{eqnarray}
or which can be proportional to $\sum|b_i|^2$ depending on the
position of zero. For the evaluation of $K$, $M_2 \approx M_{GUT}$
would be taken.

The washout effect of the asymmetry varies significantly depending
on the structure of Yukawa matrix, i.e., when the decay width is
determined by each type of Yukawa matrix listed in
Eq.(\ref{onezero}) and three more with a texture zero in
counter-position. Out of all the types of Yukawa matrices
examined, there is no such a case that Yukawa couplings originate
decays of neutrinos $N_1$ which satisfy the out-of-equilibrium
condition $K<1$ at $T=M_1$. The dilution factor $\kappa$ given in
Eq.(\ref{largek}) and Eq.(\ref{smallk}) changes in size when it is
described in terms of low energy phases $\delta$ and $\varphi$,
although the variation in dilution factor is not so remarkable as
to affect the order of magnitude as follows. If $y_{11}$ is 0,
$\kappa=(6.9-8.9)\times 10^{-3}$ as $(\delta,\varphi)$ varies
$(\pi,\pi)$ to $(0,0)$. If $y_{12}$ is 0, $\kappa = 8.0 \times 10
^{-2}$ and the value does not vary on $\delta$ and $\varphi$. If
$y_{21}$ is 0, $\kappa=(6.9-9.1)\times 10^{-2}$ as
$(\delta,\varphi)$ varies $(0,\pi)$ to $(\pi,0)$. If $y_{22}$ is
0, $\kappa=(6.5-8.7)\times 10^{-3}$ as $(\delta,\varphi)$ varies
$(0,\pi)$ to $(0,0)$. If $y_{31}$ is 0, $\kappa=(6.9-9.1)\times
10^{-2}$ as $(\delta,\varphi)$ varies $(0,0)$ to $(\pi,\pi)$. And
if $y_{32}$ is 0, $\kappa=(6.5-8.7)\times 10^{-3}$ as
$(\delta,\varphi)$ varies $(\pi,\pi)$ to $(\pi,0)$. The dilution
factor is enhanced most with the Yukawa matrix of type
$[y_{12}=0], [y_{21}=0]$ or $[y_{31}=0]$, which shows that the
amount of asymmetry survived from washout is at most 10$\%$. It is
worth reminding that the survived portion of asymmetry cannot
exceed 17$\%$ even if one took $K=0$. The rest three types of
Yukawa couplings give rise to even lower dilution factor. When
$T<M_1$, the Boltzmann equations still depict the finite value of
$\kappa$ as $M_1/T$ increases for the universe evolution
\cite{Kolb:qa}\cite{Luty:un}.

\begin{widetext}
\begin{figure}
  \begin{center}
    \begin{tabular}{cc}
      \resizebox{80mm}{!}{\includegraphics[angle=0]{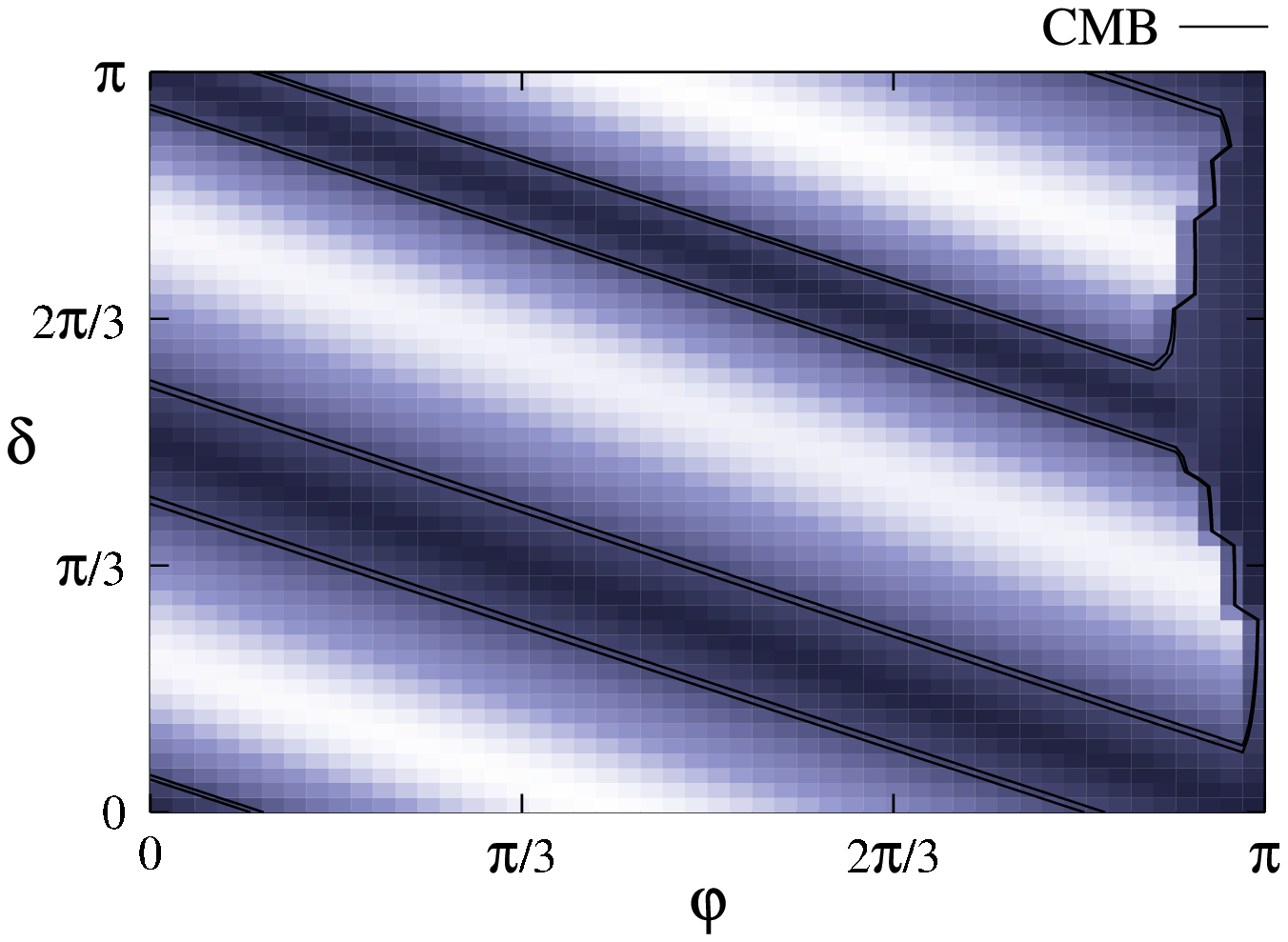}} &
      \resizebox{80mm}{!}{\includegraphics[angle=0]{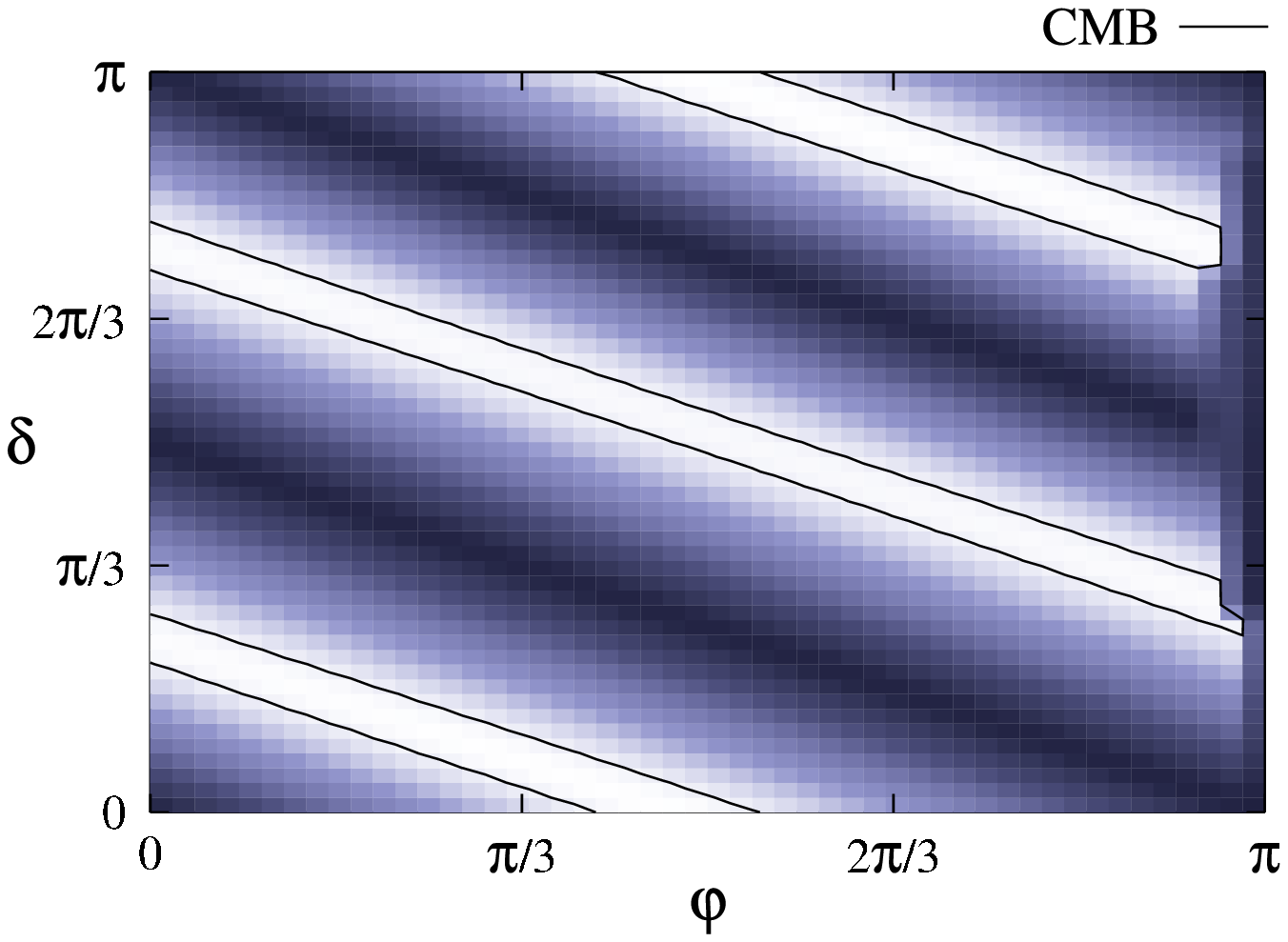}}
    \end{tabular}
    \caption{$Y_L$ vs. $\delta$ and $\varphi$ when $y_{11}=0$
            (a) $\theta=\rho$, (b) $\theta=0.6\rho$}
    \label{y110}
  \end{center}
\end{figure}
\begin{figure}
  \begin{center}
    \begin{tabular}{cc}
      \resizebox{80mm}{!}{\includegraphics[angle=0]{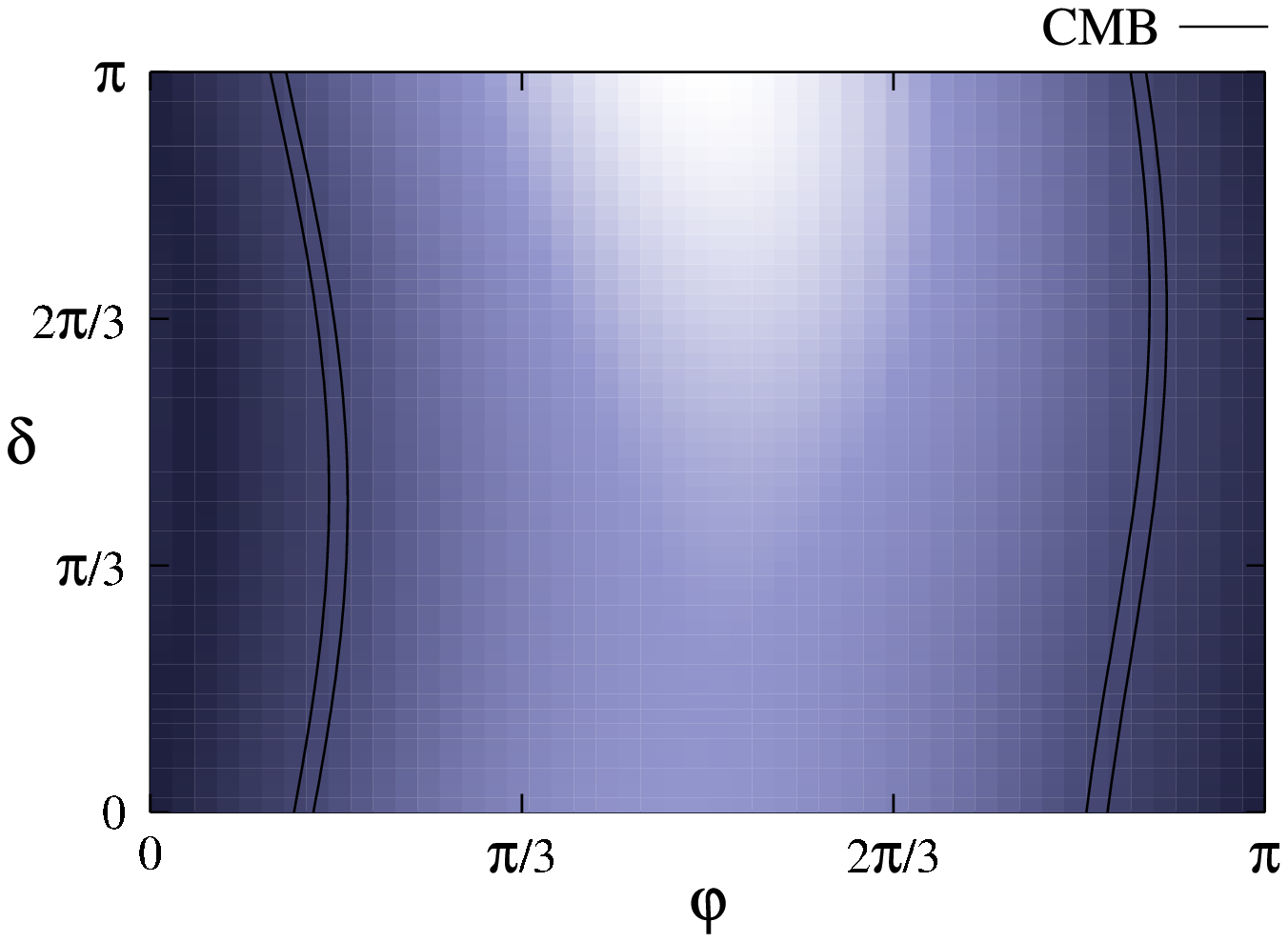}} &
      \resizebox{80mm}{!}{\includegraphics[angle=0]{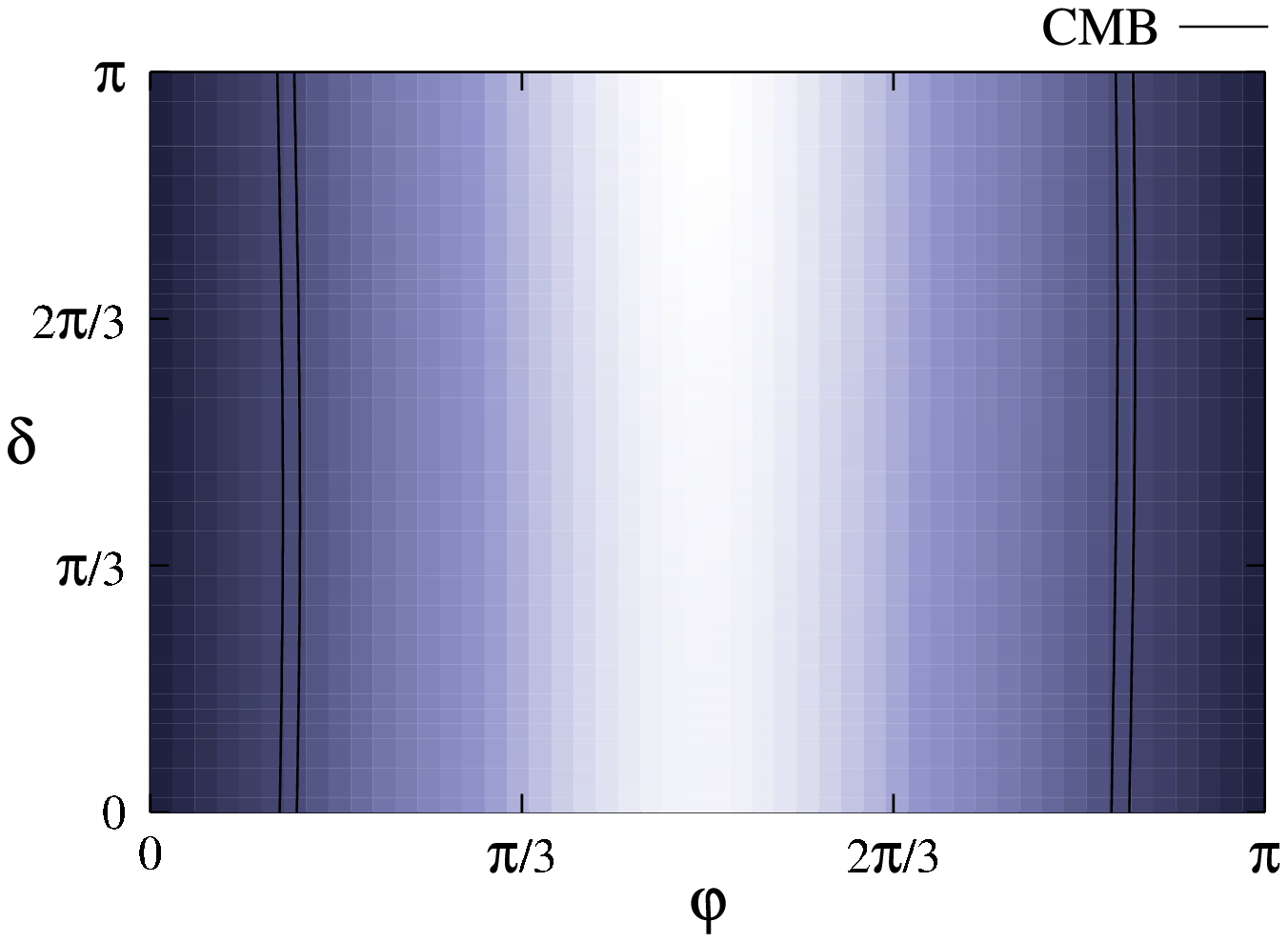}} \\
    \end{tabular}
    \caption{$Y_L$ vs. $\delta$ and $\varphi$ when $y_{22}=0$
            (a) $\theta=\rho$, (b) $\theta=0.1\rho$}
    \label{y220}
  \end{center}
\end{figure}
\end{widetext}

The amount of lepton asymmetry in Eq.(\ref{aalepto}) is now given
as a function of $\delta$ and $\varphi$ as well as $\theta_{13}$,
    \begin{eqnarray}
        Y_L =
        \frac{1}{g^*}\kappa(\delta,\varphi,\theta_{13})
        \epsilon_1(\delta,\varphi,\theta_{13}),
    \end{eqnarray}
which imbeds $\Delta_1$'s analyzed in the last section and $K$'s
in Eq.(\ref{kay}). The Fig.\ref{y110} and Fig.\ref{y220} show the
dependence of $Y_L$ on the Dirac phase $\delta$ and the Majorana
phase $\varphi$. The brighter part represents a region of higher
$Y_L$, while the darker represents that of lower $Y_L$. The
contour is the amount of $Y_L$ dictated by Cosmic Microwave
Background observation \cite{WMAP1}\cite{Harvey:1990qw}.
Hereafter, it will be denoted by $Y_L^{CMB}$. The dark region
outside of the contour $Y_L^{CMB}$ in $\delta$-$\varphi$ space is
ruled out. In Fig.1, the pattern of hue and the contour of
$Y_L^{CMB}$ manifest the proportionality of the obtained $Y_L$ to
$\sin(\varphi+2\delta)$ as analyzed in Eq.(\ref{imaginary1}).

\begin{widetext}
\begin{figure}
  \begin{center}
    \begin{tabular}{cc}
      \resizebox{80mm}{!}{\includegraphics[angle=0]{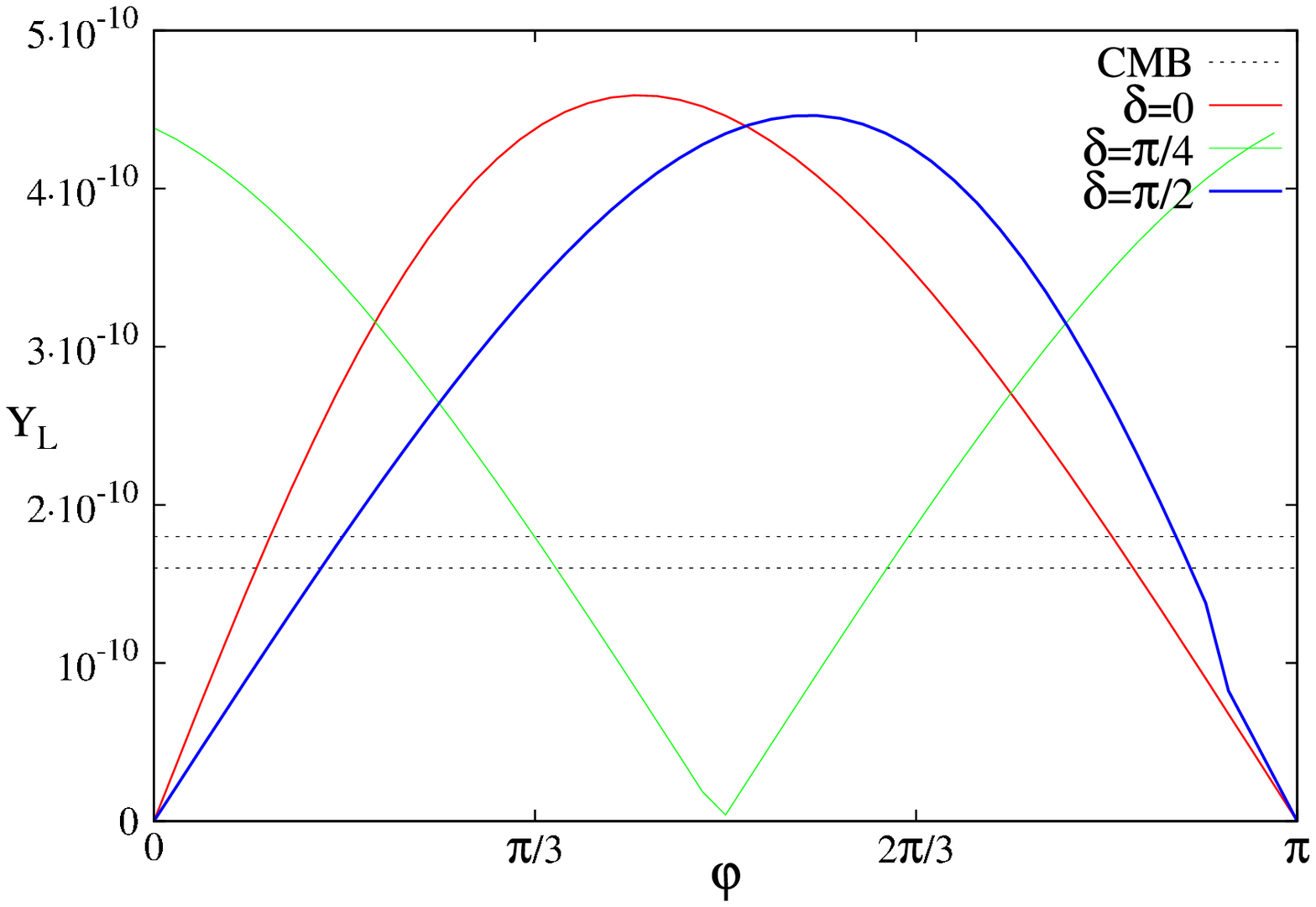}}
      \resizebox{80mm}{!}{\includegraphics[angle=0]{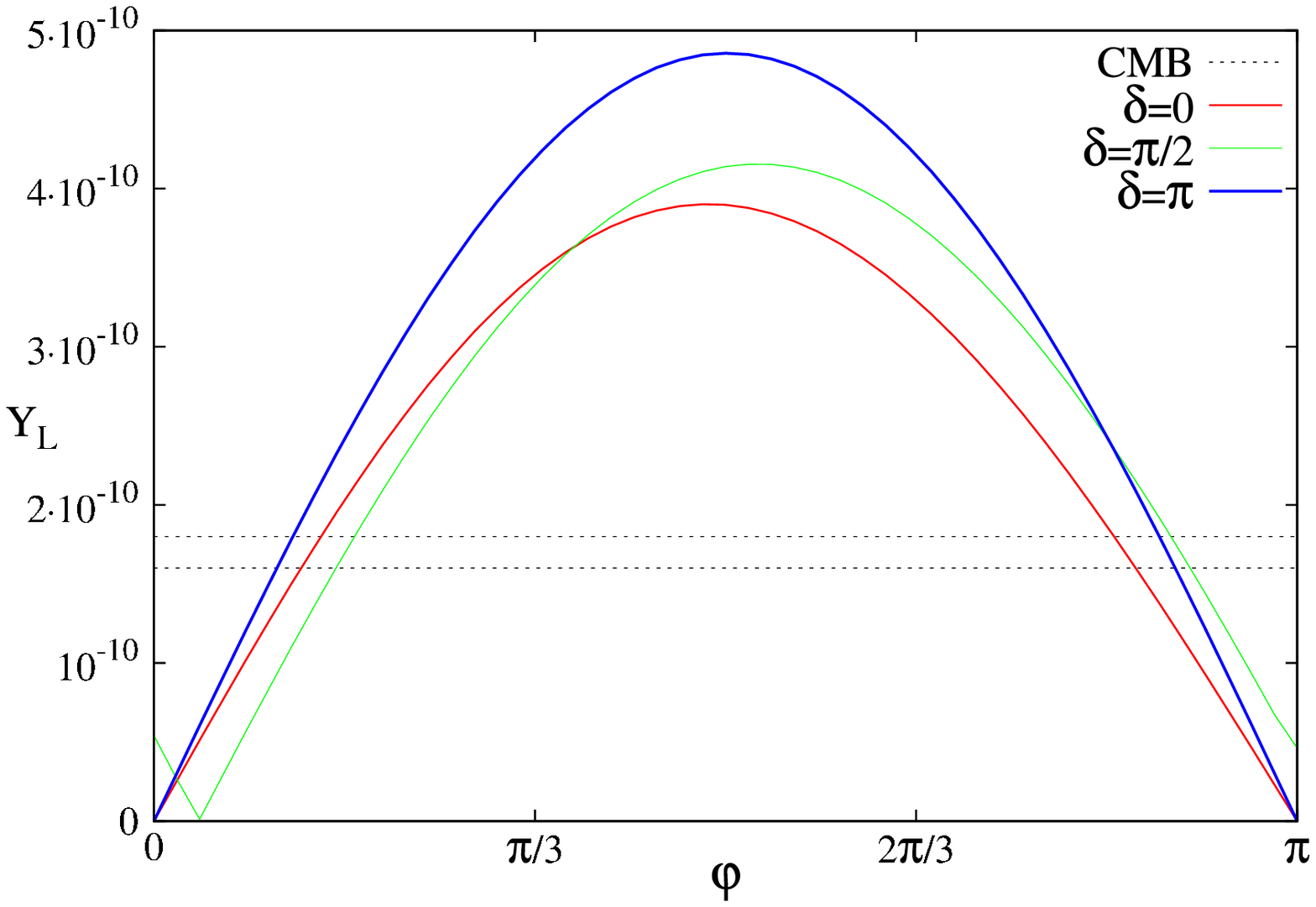}} \\
    \end{tabular}
    \caption{(color online) $Y_L$ vs. $\varphi$ at $\theta=\rho$,
    the cross sections
        by particular values of $\delta$'s of the curves (a) in
        Fig.1-(a) for $y_{11}=0$, and (b) in Fig.2-(a) for $y_{22}=0$.}
    \label{xsections}
  \end{center}
\end{figure}
\end{widetext}

When $\theta=\rho$ as in Fig.\ref{y110} (a), the sufficient amount
of lepton asymmetry prevails in most part of $\delta$-$\varphi$
space. When $\theta=0.6\rho$ as in Fig.\ref{y110} (b), only
strongly restricted region just above $Y_L^{CMB}$ can be regarded
as plausible for leptogenesis. In Fig.\ref{y220} for $y_{21}=0$,
it shows clearly that the size of angle $\theta_{13}$ does not
affect the amount of the lepton asymmetry. The shape of the hue in
Fig.\ref{y110} and Fig.\ref{y220} can be shown by their
cross-sections as in Fig.\ref{xsections}. In Fig.\ref{xsections}
(a) with $y_{11}=0$, the Majorana phase is strongly constrained by
the Dirac phase, while the $Y_L$ vs. $\varphi$ relation does not
rely on $\delta$ so significantly in Fig.\ref{xsections} (b) with
$y_{22}=0$ as in Fig.\ref{xsections} (a). Aspects of the relation
of lepton asymmetry and observable CP phases for the case with
$y_{32}=0$ are similar to those for the case with $y_{22}=0$,
regarding the amount of the asymmetry and the correlation of two
types of phases.

For the other three cases with $y_{12}=0,~y_{21}=0$ and
$y_{31}=0$, the amount of lepton asymmetry is enhanced by about
two orders of magnitude since washout effect is suppressed and the
size of CP asymmetry is increased in comparison with those of the
previous three cases. From Fig.\ref{xsections}, it is obvious that
most range in $\delta$-$\varphi$ space is involved with the amount
of the asymmetry above $Y_L^{CMB}$. As for the correlation of two
CP phases, the $\delta$ and $\varphi$ with $y_{12}=0$ also
constrain each other by $\sin(\varphi+2\delta)$ as so do the
phases with $y_{11}=0$. When $y_{21}=0$ or $y_{31}=0$, the
asymmetry mainly depends on $\sin\varphi$ as seen for $y_{22}=0$
or $y_{32}=0$.

It has been revealed that a Yukawa matrix with a texture zero at
$y_{11}=0$ or $y_{12}=0$ originates the lepton asymmetry strongly
correlated with $\theta_{13}$ and $\delta$. The next section
includes the discussion on the possible amount of the lepton
asymmetry to the size of CP violation in neutrino oscillation and
the amplitude of neutrinoless double beta decay.

\section{Remarks}

\begin{widetext}
\begin{figure}
  \begin{center}
    \begin{tabular}{cc}
      \resizebox{80mm}{!}{\includegraphics[angle=0]{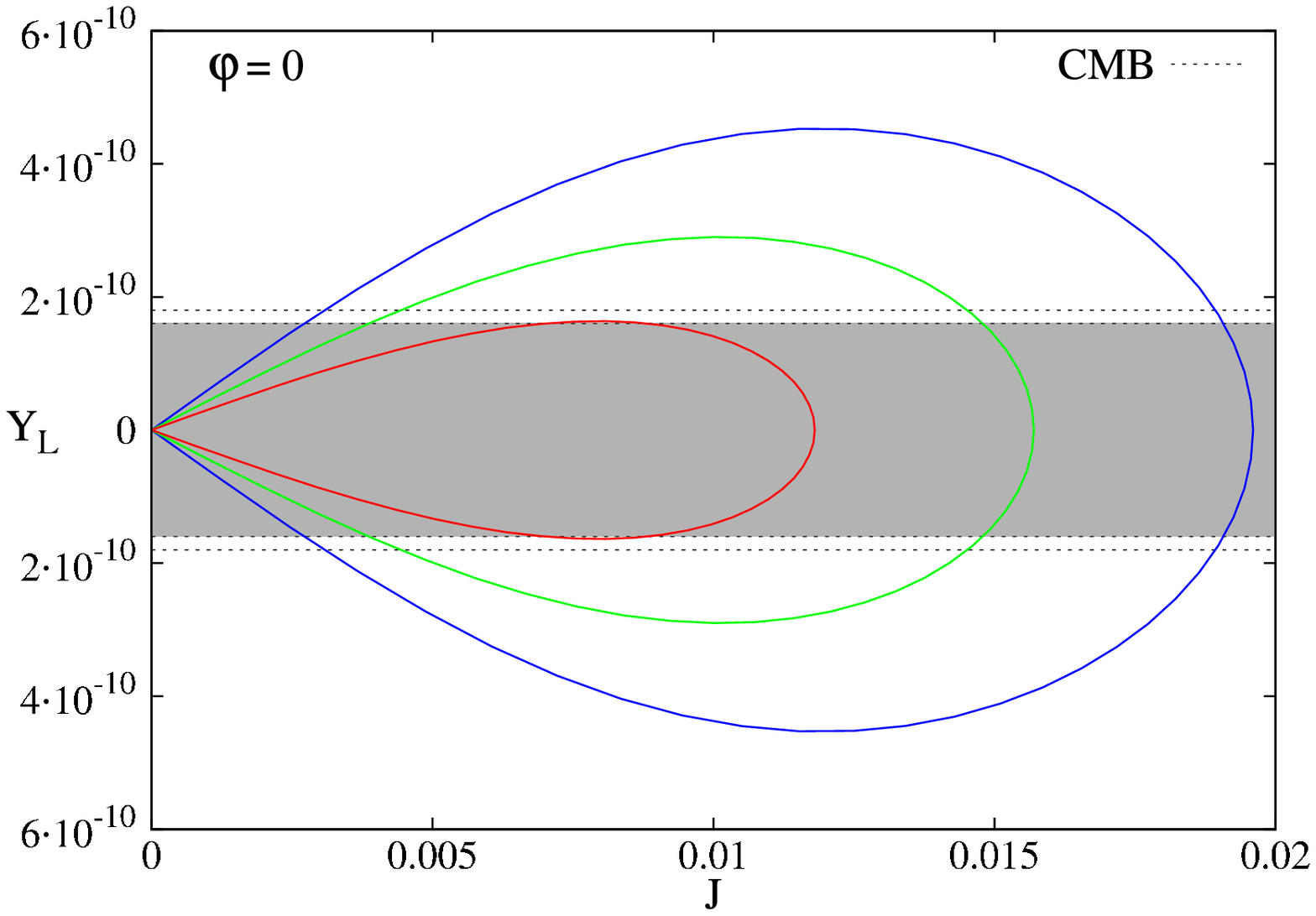}} &
      \resizebox{80mm}{!}{\includegraphics[angle=0]{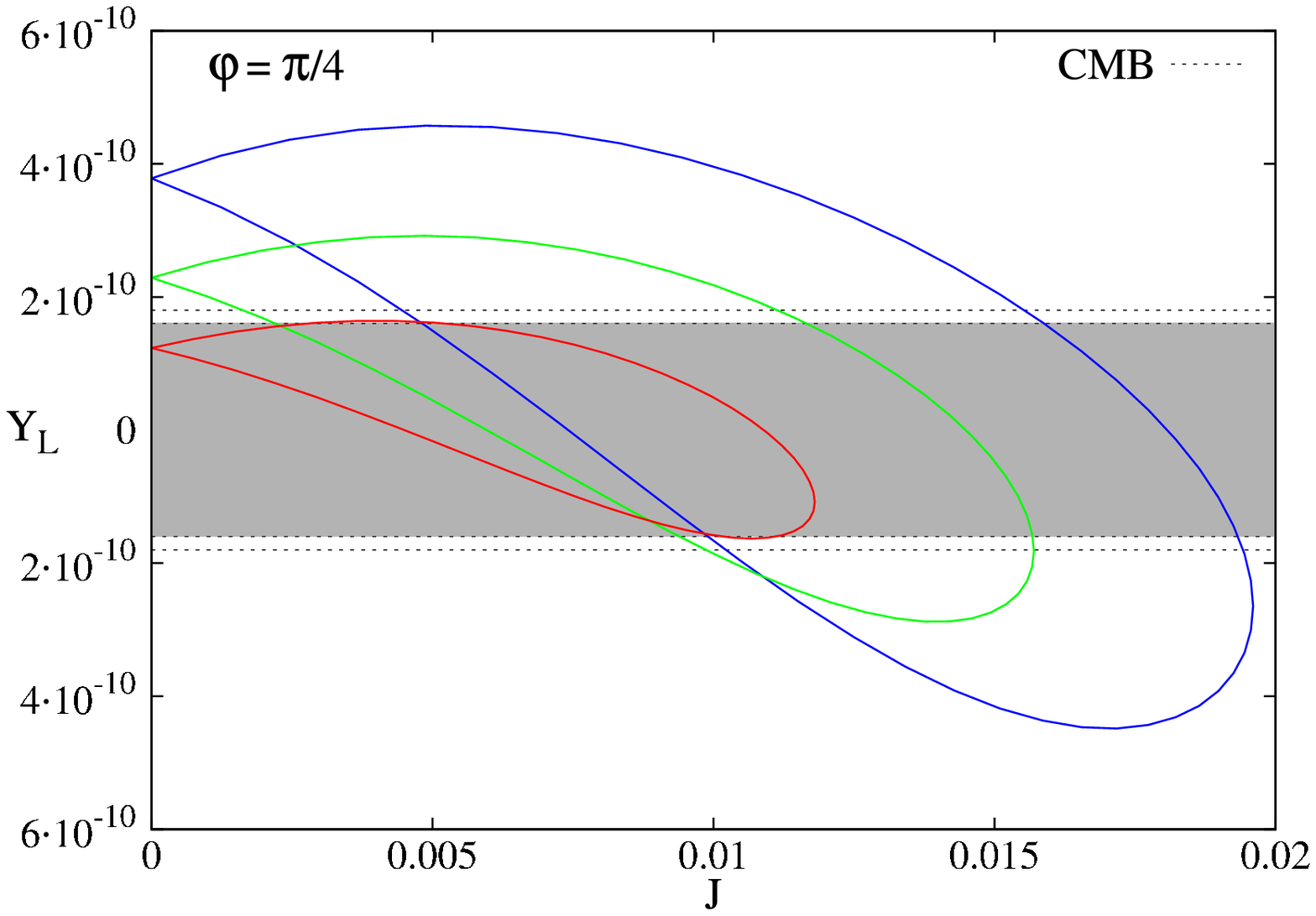}} \\
      \resizebox{80mm}{!}{\includegraphics[angle=0]{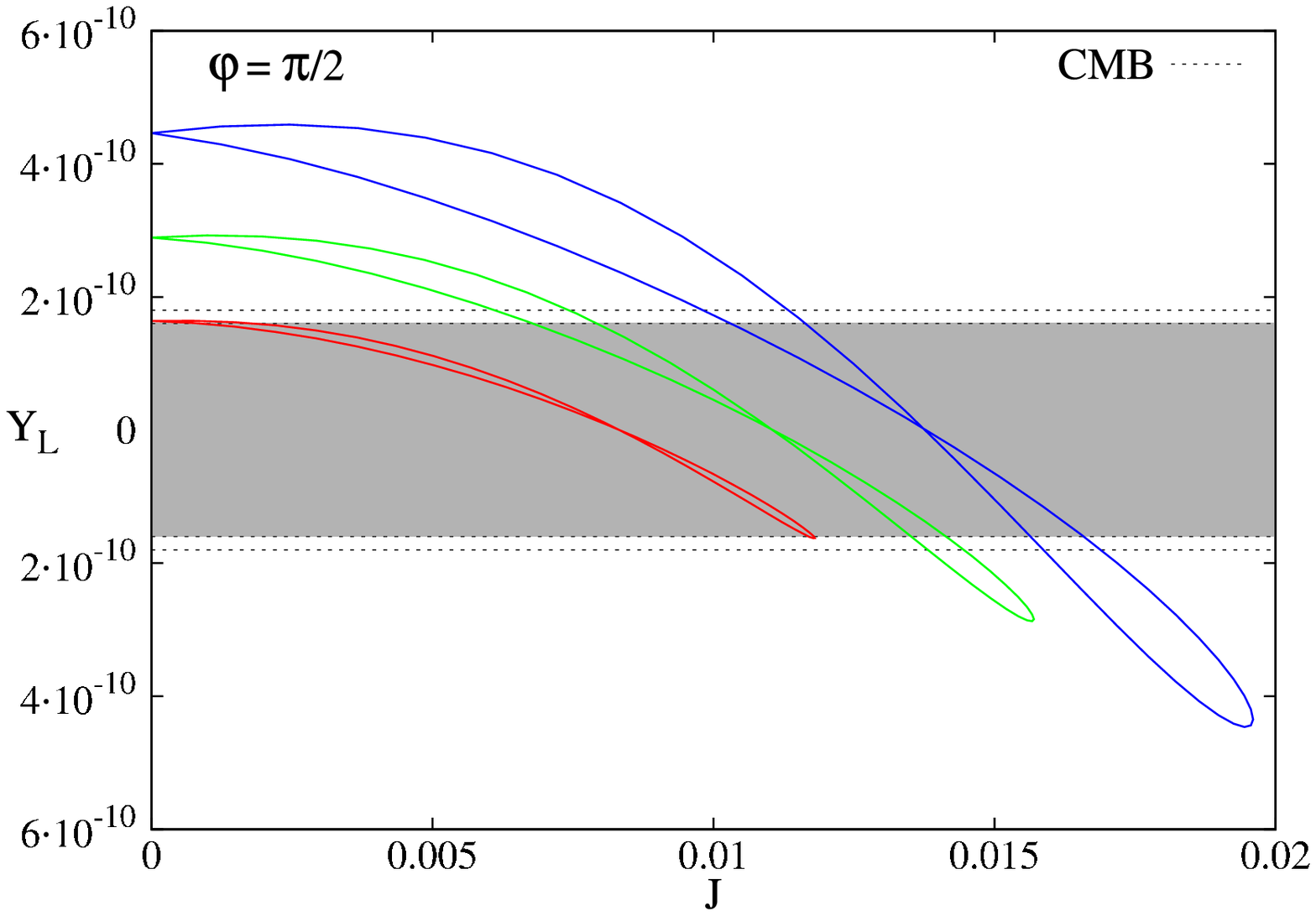}} &
      \resizebox{80mm}{!}{\includegraphics[angle=0]{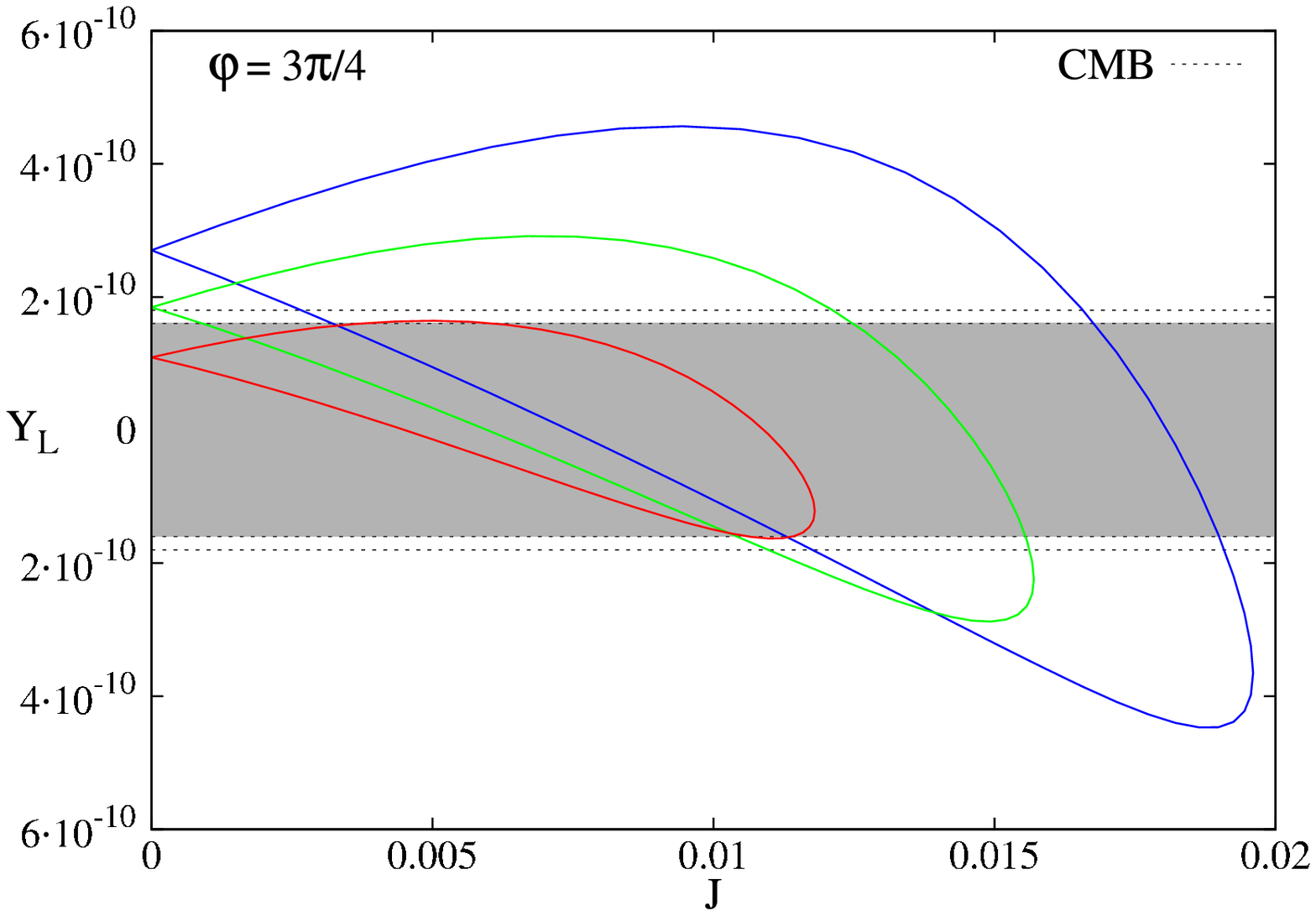}} \\
    \end{tabular}
    \caption{(color online)\hspace{3pt} $Y_L$ vs. $\mathcal{J}$
    for chosen values of Majorana phase.~
    The large(blue), medium(green), and
        small(red) curves correspond to
        $\sin^2{2\theta_{13}} = 0.0255, ~0.0164,$ and $0.00922$,
        respectively.}
    \label{YLvsJ}
  \end{center}
\end{figure}
\end{widetext}

\begin{widetext}
\begin{figure}
  \begin{center}
    \begin{tabular}{cc}
      \resizebox{76mm}{!}{\includegraphics[angle=0]{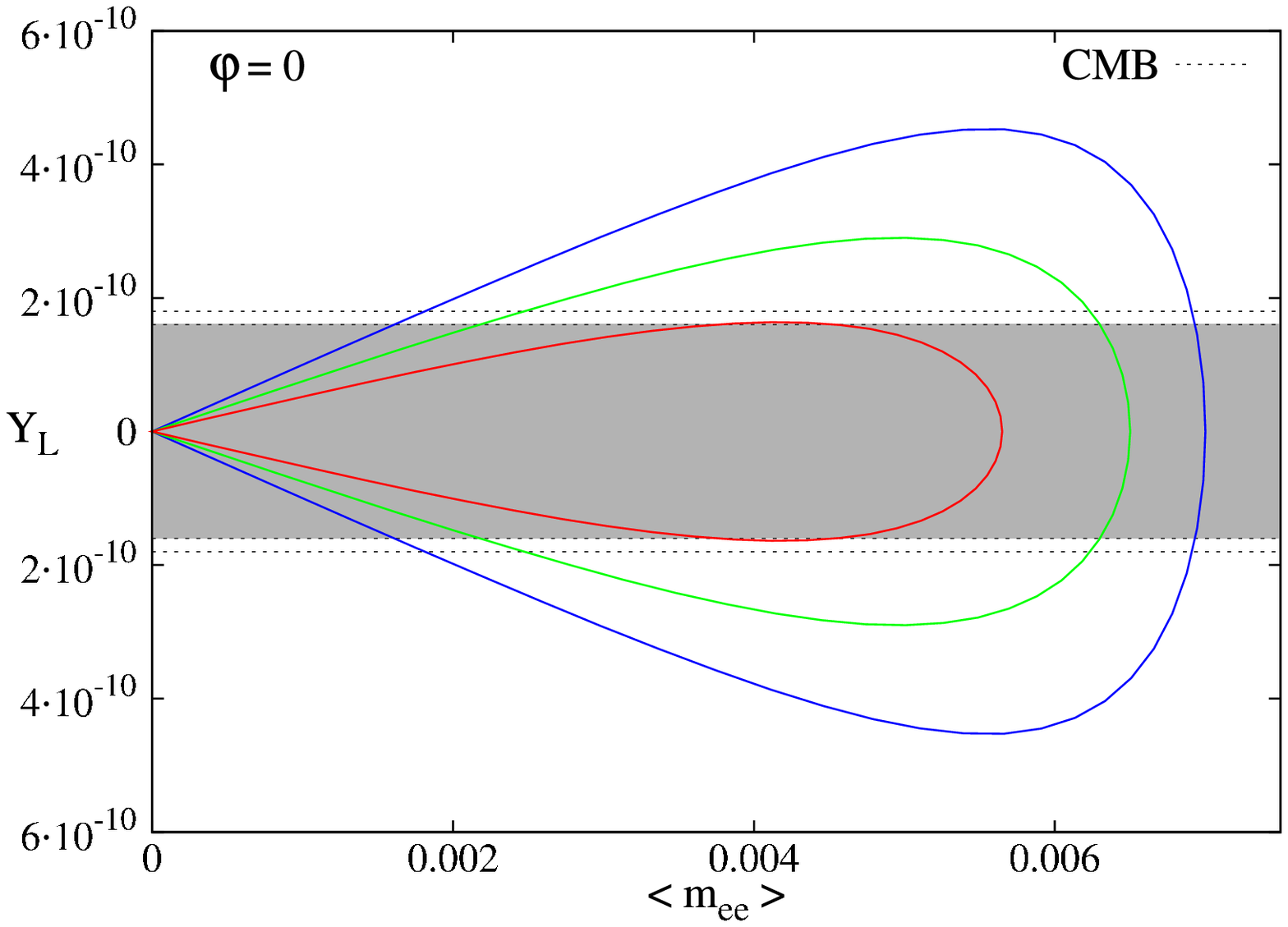}} &
      \resizebox{76mm}{!}{\includegraphics[angle=0]{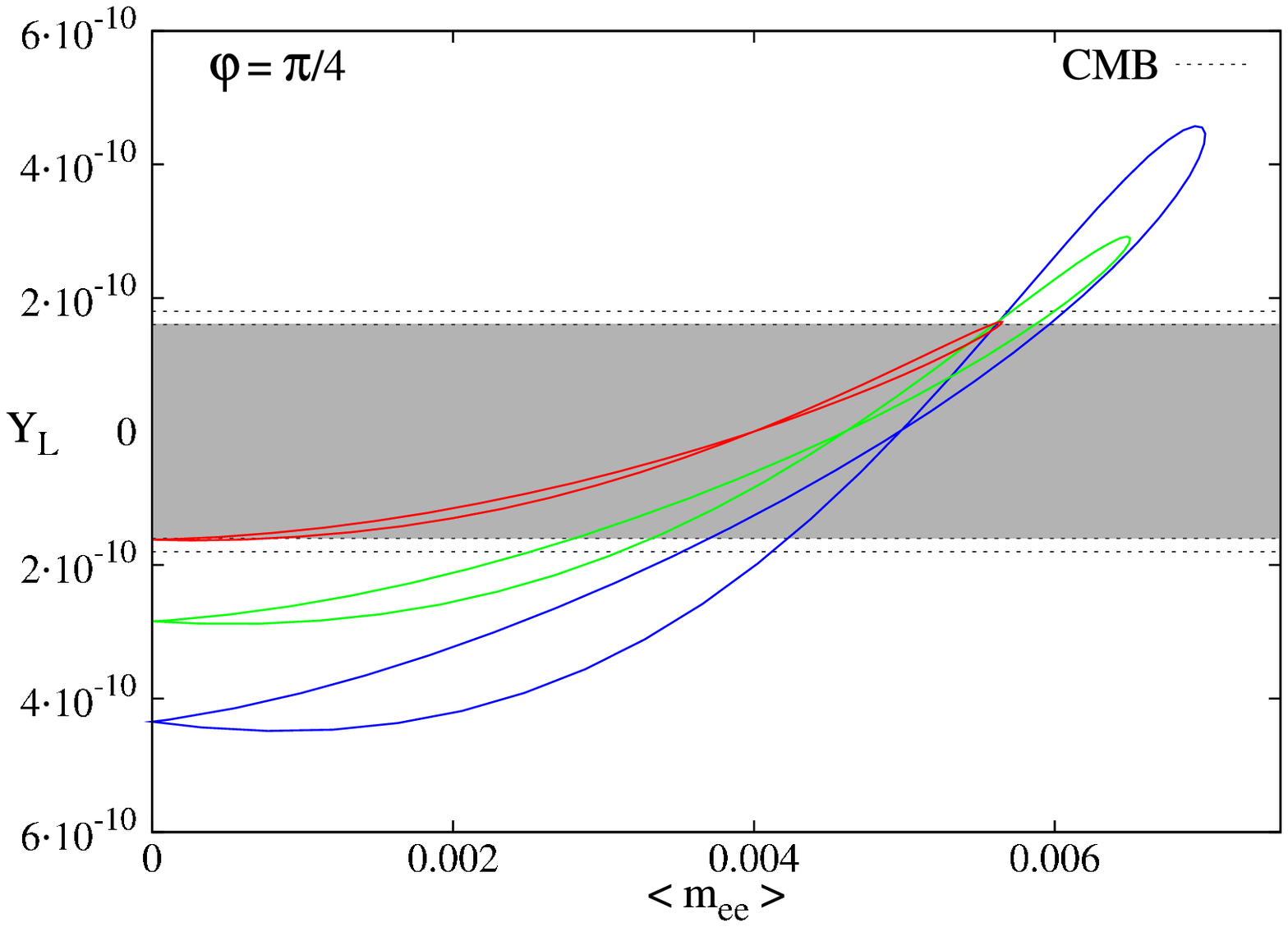}} \\
      \resizebox{76mm}{!}{\includegraphics[angle=0]{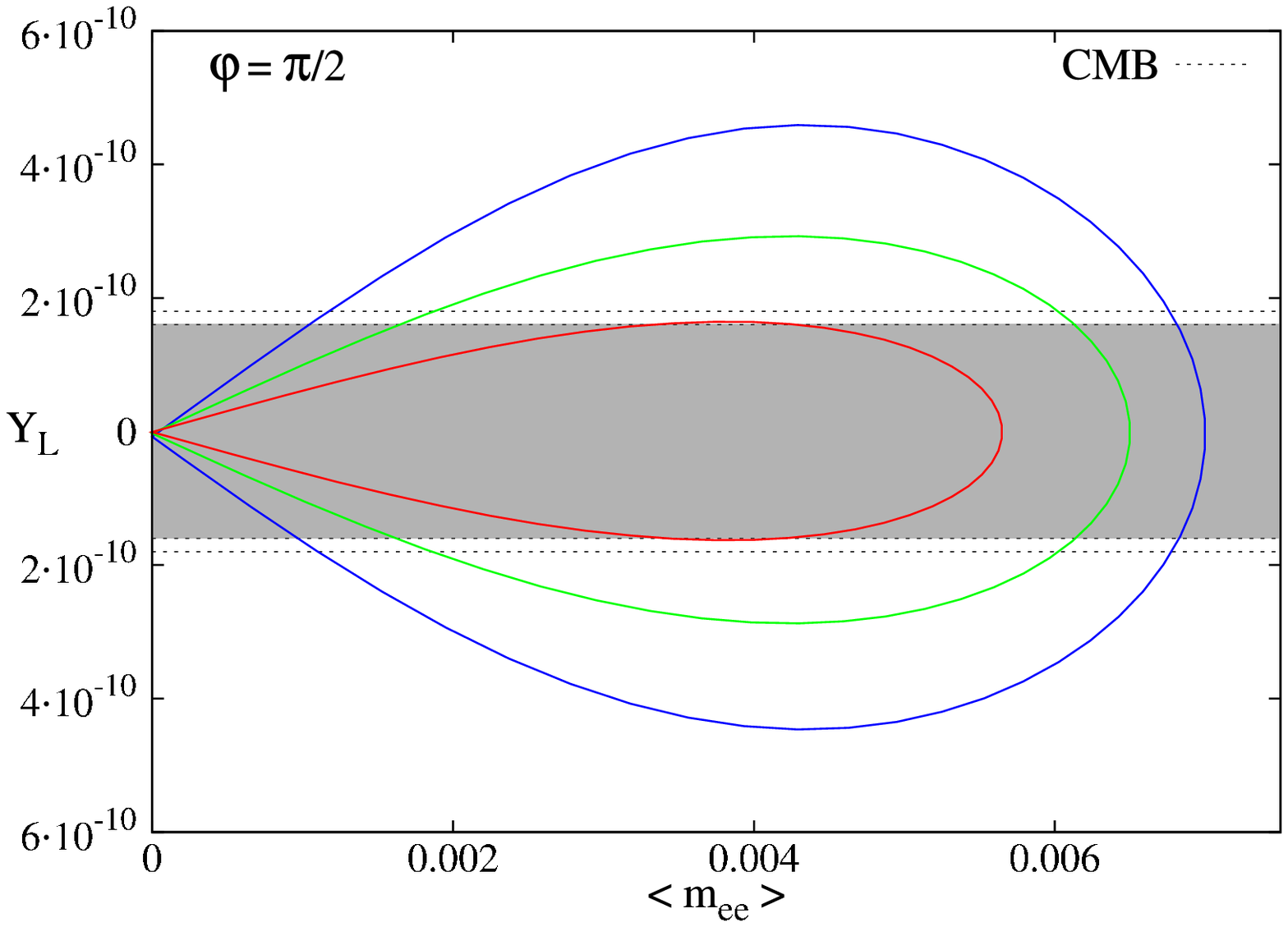}} &
      \resizebox{76mm}{!}{\includegraphics[angle=0]{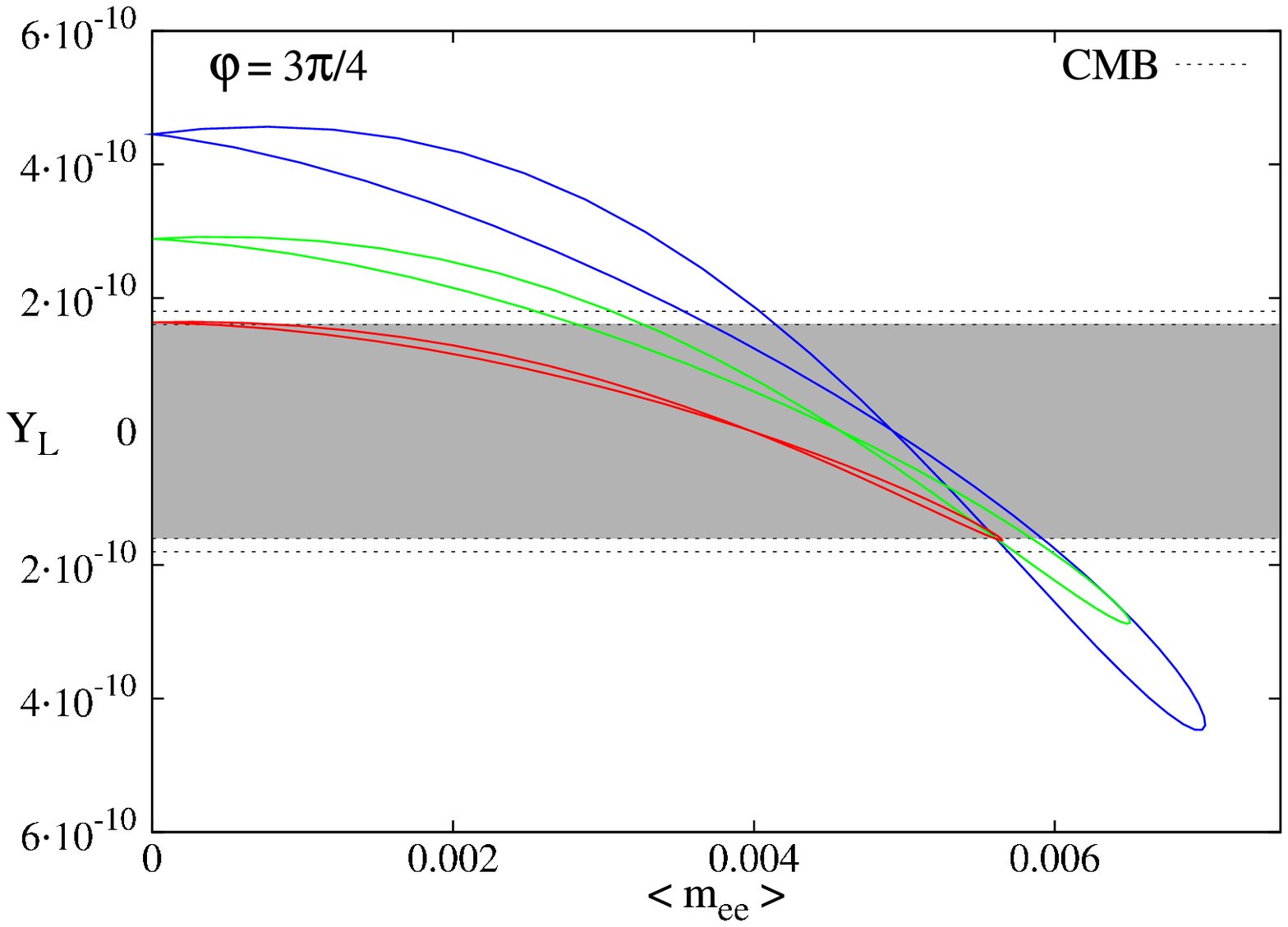}} \\
    \end{tabular}
    \caption{(color online)\hspace{3pt} $Y_L$ vs. $<m_{ee}>$
    for chosen values of Majorana phase.~
    The large(blue), medium(green), and
        small(red) curves correspond to
        $\sin^2{2\theta_{13}} = 0.0255, ~0.0164,$ and $0.00922$,
        respectively.}
    \label{YLvsMee}
  \end{center}
\end{figure}
\end{widetext}

There has been consideration about the way the amount of the
lepton asymmetry is dependent on the low energy measurable
$\theta_{13}, ~\varphi,$ and $ \delta$ in Eq.(\ref{morder}) and
how the dependence varies according to structures of Yukawa
matrices, even though those different types of Yukawa matrices can
cause the same prediction for the low energy parameters. In
Fig.\ref{YLvsJ} and Fig.\ref{YLvsMee}, the case with Yukawa matrix
with a texture zero at $y_{11}=0$ is considered for the
compatibility of the production of sufficient lepton asymmetry and
the possible measurement of Jarlskog invariant of $\mathcal{J}$
and the amplitude of neutrinoless double beta decay. For the
particular case, the figures clearly show that if
$\sin^2{2\theta_{13}}$ is smaller than 0.01, the lepton asymmetry
obtained from the Yukawa couplings is below the amount of
asymmetry based on the baryon asymmetry observation.

If $\rho$ and $\theta$ in Eq.(\ref{morder}) has such relations as
$\theta=\rho, ~\theta=0.8\rho,$ and $\theta=0.6\rho$, each case
corresponds to $\sin^2{2\theta_{13}} = 0.0255, ~0.0164,$ and
$0.00922$. Although the 5 elements in Eq.(\ref{key}) were
determined by taking a texture zero among themselves, the relative
sign of two columns is not yet specified and, accordingly, only
the absolute value of the asymmetry is concerned with the Jarlskog
invariant as shown in Fig.\ref{YLvsJ}.

Once the hierarchy in masses of the right-handed heavy neutrinos
is determined, the lepton asymmetry from Yukawa couplings with a
texture zero at other than $y_{11}$ and $y_{12}$ has main
correlation with the Majorana phase $\varphi$. The correlation of
$Y_L$ to the small mixing angle $\theta_{13}$ and the Dirac phase
$\delta$ appears strongly when $y_{11}$ or $y_{12}$ is zeroed. In
other words, the Dirac phase $\delta$ and the Majorana phase
$\varphi$ constrains each other so that only a certain region of
$\delta$-$\varphi$ space can be compatible with the leptogenesis.

\vspace{10pt}
\begin{acknowledgments}
This research was supported by the Chung-Ang University Research
Grants in 2004.
\end{acknowledgments}

\end{document}